\documentclass[a4paper,12pt]{article}
\usepackage{amsthm}
\usepackage{amssymb}
\usepackage{amsmath}
\usepackage{amsfonts}
\usepackage{color}
\usepackage{graphicx}
\usepackage[square, comma, sort&compress, numbers]{natbib}

\newtheorem{remark}{Remark}[section]

\newtheorem{lemma}{Lemma}[section]
\newtheorem{theorem}{Theorem}[section]

\newtheorem{definition}{Definition}[section]

\def\b1{\mbox{\boldmath $1$}}

\newenvironment{demo*}{\vspace{3mm}\noindent{\bf Proof.}}{\hfill $\Box$ \vspace{3mm}}

\begin{document}
\title{\bf \Large {Sharp convex bounds on the  aggregate sums--An alternative  proof }}
{\color{red}{\author{
\normalsize{Chuancun Yin\;\; Dan Zhu}\\
{\normalsize\it  School of Statistics,  Qufu Normal University}\\
\noindent{\normalsize\it Shandong 273165, China}\\
e-mail:  ccyin@mail.qfnu.edu.cn}}}
\maketitle
\vskip0.01cm
\noindent{\large {\bf ABSTRACT}}  { It is well known that  a random vector with given marginal distributions is comonotonic if and only if it has the
largest sum with respect to the convex order [ Kaas,  Dhaene,  Vyncke,  Goovaerts, Denuit (2002), A simple geometric
proof that comonotonic risks have the convex-largest sum, ASTIN Bulletin 32, 71-80.  Cheung (2010), Characterizing a comonotonic random vector by the distribution of the sum of its components, Insurance: Mathematics and Economics 47(2), 130-136] and that  a random vector with given marginal distributions   is mutually exclusive if and only if it has the minimal convex sum [Cheung and Lo (2014), Characterizing mutual exclusivity as the strongest negative
multivariate dependence structure, Insurance: Mathematics and Economics 55, 180-190].  In this
note, we give a new  proof of these two results  using the theories of distortion risk measure and  expected utility. }

\medskip

{Keywords:}  {\rm  {{Comonotonicity; Convex order;  Distortion risk measure;    Mutual exclusivity; Stop-loss order}} }


\numberwithin{equation}{section}
\section{Introduction}\label{intro}

After years of efforts made by  researchers, the study of sharp convex bounds on the sum of random
variables (also known as aggregate sums) with given marginal
distributions but unknown dependence structure  has   achieved a lot of significant results. Mathematically, given an arbitrary Fr\'echet space
${\cal R}(F_1, \cdots,  F_n)$ of all random vectors having $F_1,\cdots, F_n$ as marginal distributions, the aim is to find two random vectors $(X_1^m,\cdots,X_n^m)$
and $(X_1^M,\cdots,X_n^M)$ belonging to ${\cal R}(F_1, \cdots,  F_n)$   such that
$$\sum_{i=1}^n X_i^m\le_{cx}\sum_{i=1}^n X_i\le_{cx}\sum_{i=1}^n X_i^M$$
for any $(X_1,\cdots,X_n)\in {\cal R}(F_1, \cdots,  F_n)$, where $\le_{cx}$ denotes the convex order. By definition,
  for a pair of random variables $X$ and $Y$, we say that $X$ is less that $Y$ in the sense of convex order, denoted as $X\le_{cx}Y$, if $Ef(X)\le Ef(Y)$ for every convex function $f$, provided that expectations $Ef(X)$ and  $Ef(Y)$ exist. In
actuarial science, it is common to define convex order by using a stop-loss transform:
  $X\le_{cx}Y \Leftrightarrow EX=EY$ and   $X\le_{sl} Y$.
  Here $X$ is said to precede $Y$ in the stop-loss order sense, notation $X\le_{sl} Y$, if and only if $X$ has lower stop-loss premiums than $Y$:
  $$ E(X-d)_+\le E(Y-d)_+, \;\;-\infty<d<\infty.$$
A summary of other characterizations and properties of convex order can be found e.g. in Denuit et al. (2005),  Shaked and Shanthikumar
(2007).

Comonotonicity plays a crucial role  in determining   convex upper bound  on aggregate sum. Let us recall the definition.
For any ${\bf X}\in {\cal{R}}(F_1, \cdots, F_n)$, ${\bf X}$
is said to be comonotonic if
$$F_{\bf X}({\bf x})=\min_{1\le k\le n}F_k(x_k),\ \forall\  {\bf x}=(x_1,x_2,\cdots, x_n)\in \Bbb{R}^n. $$
Equivalently, ${\bf X}$ is  comonotonic if and only if ${\bf X}\stackrel{d}{=}(F_1^{-1}(U),\cdots, F_n^{-1}(U))$, where $U$ is a random variable uniformly distributed on the interval $[0,1]$,
denoted as $U\sim U[0,1]$.
The concept of comonotonicity was introduced by Yaari (1987) and Schmeidler
(1986). For more details and  other
characterizations about the concept of comonotonicity
and its applications in actuarial science and finance, we refer to the
overview papers by Dhaene et al.(2002a, 2002b) and more recently in
Deelstra et al. (2010).
Let $S$ be the sum $X_1+\cdots+X_n$ and $S^c$ be the comonotonic sum $X_1^c+\cdots+X_n^c$, where $(X_1^c,\cdots,X_n^c)$ is the comonotonic counterpart of ${\bf X}=(X_1,\cdots,X_n)$.   A well-known result  between the sums $S$ and $S^c$ says that $S\le_{cx} S^c$. Proofs of this fundamental result
in the bivariate case can be found in Dhaene and Goovaerts (1996,
1997). M\"uller (1997) extended the result to higher dimensions  as a special case of the concept
of supermodular ordering.
 A simple geometric argument is given in Kaas et al. (2002) and Cheung (2010a) provided a new proof using the theory of majorization. The converse remains valid under the assumption that  all marginal distribution functions are continuous and that the underlying probability space $(\Omega, \cal{F}, P)$ is atomless;  see Cheung (2008).   This continuity
assumption on the marginals was removed by Cheung (2010b).  A new and simple proof  without the assumption that the underlying probability
space $(\Omega, \cal{F}, P)$  is atomless was given by Mao and  Hu (2011).  To summarize above results   we arrive at  the following theorem.
\begin{theorem} If $(X_1^*,\cdots, X_n^*)\in{\cal{R}}(F_1, \cdots, F_n)$, then  $(X_1^*,\cdots, X_n^*)$ is comonotonic if, and only if $$X_1+\cdots+X_n\le_{cx} X_1^*+\cdots+X_n^* \,\, {\rm for\ all}\,\, (X_1,\cdots, X_n)\in {\cal{R}}(F_1, \cdots, F_n).$$
\end{theorem}
Now we focus on  the lower convex bound of ${\cal{R}}(F_1, \cdots, F_n)$.  When  $n = 2$, the minimum sharp bound is obtained by the counter-monotonic scenario:
 $$F_1^{-1}(U)+F_2^{-1}(1-U)\le_{cx}X_1+X_2\ {\rm for\ any} \    (X_1, X_2)\in{\cal{R}}(F_1, F_2),$$
 where  $U$ is a random variable uniformly distributed on the interval $[0,1]$; see, for example,  Denuit et al. (2005, p. 290). Moreover, Cheung and Lo (2013a) shows that the converse remains valid.
 However, the sharp lower convex bound for $n\ge 3$ is missing in the literature due to the fact that counter-monotonicity cannot be generalized to $n\ge 3$ without losing its minimality with respect to convex order.
  In a special case, when $F_1,\cdots,F_n$ are on ${\Bbb R}^+$ with $\sum_{i=1}^n (1-F_i(0))\le 1$,  the convex lower  bound is obtained by the mutually exclusive scenario:
  $$X^*_1 +\cdots+X^*_n \le _{cx} X_1+\cdots+X_n$$
   for any $(X_1, \cdots, X_n) \in  {\cal{R}}(F_1, \cdots, F_n)$, where   $(X^*_1, \cdots, X^*_n) \in  {\cal{R}}(F_1, \cdots, F_n)$ and $P(X_i^*>0, X_j^*>0)=0$ for all $i\neq j$;  see Dhaene and Denuit (1999, Theorem 10).
 When the marginals $F_1,F_2,\cdots,F_n$ are two-point distributions, the result can be found in Hu and Wang (1999).
   Mutual exclusivity can be considered as the strongest negative
dependence structure in a multivariate setting. It was first studied
in Dhaene and Denuit (1999), and recently revisited, generalized
and further characterized  in Cheung and Lo (2014).
\begin{definition} (Cheung and Lo (2014))\ Let $X_1, \cdots, X_n$ be random variables with essential infima
$l_1,\cdots, l_n$ and essential suprema $u_1, \cdots, u_n$ respectively. They   are said to be\\
(i) mutually exclusive from below if $P(X_i > l_i, X_j > l_j) = 0$ for all $i \neq j$;\\
(ii) mutually exclusive from above if $P(X_i < u_i, X_j < u_j) = 0$ for all $i \neq j$.
\end{definition}
 The following theorem, due to Cheung
and Lo (2014),  concerning mutually exclusive random
variables and the  minimal lower bound in convex order.
\begin{theorem}  (Cheung and Lo (2014)) \ Let ${\bf X}^*=(X^*_1,\cdots, X^*_n)$ be a fixed random
vector in ${\cal{R}}(F_1, \cdots, F_n)$  ($n\ge 3$) which satisfies $\sum_{i=1}^n (1-F_i(l_i))\le 1$ or $\sum_{i=1}^n F_i(u_i-)\le 1$. Then  ${\bf X}^*$    is mutually exclusive  if and only if
$$ X_1^*+\cdots+X_n^*\le_{cx}  X_1+\cdots+X_n$$
 for all $(X_1, \cdots, X_n) \in  {\cal{R}}(F_1, \cdots, F_n)$.
\end{theorem}
In this short note, we give a new  proof of Theorems 1.1 and 1.2.  The proof is given in the next two sections.

 \vskip 0.2cm
 \section{A new proof of Theorem 1.1}
\setcounter{equation}{0}

To prove  Theorem 1.1, we need two useful lemmas. Here are some notations.
Let $F_X$ be the cumulative distribution function of random variable $X$ and the decumulative distribution function
is denoted by $\bar{F}_X$, i.e. $\bar{F}_X(x)=1-F_X(x)=P(X>x)$. A distortion function is defined as a non-decreasing
function $g : [0, 1] \rightarrow [0, 1]$ such that $g (0)=0$ and $g(1)=1$. The distortion
risk measure associated with distortion function $g$ is denoted by $\rho_{g}[\cdot]$ and
is defined by
 $$\rho_{g}[X]=\int_0^{+\infty}g(\bar{F}_X(x))dx+\int_{-\infty}^0 [g(\bar{F}_X(x))-1]dx,$$
 for any random variable $X$,
provided at least one of the two integrals above is finite. If $X$ is a non-negative  random variable, then $\rho_{g}$ reduces to
$$\rho_{g}[X]=\int_0^{+\infty}g(\bar{F}_X(x))dx.$$
 Obviously, a concave distortion function is continuous on $(0,1]$ and can only jump at $0$. In view of Dhaene et al. (2012, Theorem 6) we know  that for any concave distortion function  $g$,   one can rewrite  $\rho_{g}[X]$  as
$$\rho_{g}[X]=\int_{[0,1]}VaR_{1-q}[X]dg(q),$$
where
$VaR_p[X]=\inf\{x|F_X(x)\ge p\}$.

The following theorem shows that stop-loss order can be
characterized in terms of ordered concave distortion risk measures; see  Dhaene et al. (2000) and Dhaene et al. (2006). Here we provide a short proof.
\begin{lemma}
For any random pair $(X , Y )$ we have that $X\le_{sl} Y$ if and only if their respective
concave distortion risk measures are ordered:
$X\le_{sl} Y  \Leftrightarrow  \rho_{g}[X]\le \rho_{g}[Y]$ for all concave distortion functions $g$.
In particular, if E[X]=E[Y], then
$X\le_{cx} Y  \Leftrightarrow  \rho_{g}[X]\le \rho_{g}[Y]$ for all concave distortion functions $g$.
\end{lemma}
{\bf Proof}\;  For any concave distortion function $g$,  $\rho_{g}$ can be written as
$$\rho_{g}[X]=\int_0^1 TVaR_p[X]d\mu(p),$$
where $\mu$  is a probability measure and $TVaR_p$ is the tail value-at -risk at level $p$:
$$TVaR_p[X]=\frac{1}{1-p}\int_p^1 VaR_w[X]dw,$$
 which is a distortion risk measure
  corresponding to the concave distortion function
 $$g(x)=\min\left\{\frac{x}{1-p}, 1\right\}, 0<p<1.$$
 The result follows£¬ as
  $X\le_{sl} Y  \Leftrightarrow  TVaR_p[X]\le TVaR_p[Y]$ for all $p\in (0,1)$ (see Theorem 3.2 in Dhaene et al. (2006)).

The following subadditivity theorem  can be found in Dhaene et al. (2000),    the bivariate case can be found in Denneberg (1994), see also
 Wang and Dhaene (1998).
\begin{lemma}
For any concave distortion function $g$ and  $(X_1,\cdots, X_n)\in{\cal{R}}(F_1, \cdots, F_n)$, we have
$$\rho_{g}[X_1 +\cdots+ X_n]\le \rho_{g}[X_1] +\cdots+\rho_{g}[X_n].$$
\end{lemma}

{\bf Proof of Theorem 1.1}\ First we assume $(X_1^*,\cdots, X_n^*)\in{\cal{R}}(F_1, \cdots, F_n)$ is comonotonic. For any concave distortion function $g$ and $(X_1,\cdots, X_n)\in {\cal{R}}(F_1, \cdots, F_n)$, by Lemma 2.2   we have
\begin{eqnarray}
\rho_{g}[X_1 + \cdots+X_n]\le \rho_{g}[X_1]+\cdots+\rho_{g}[X_n].
\end{eqnarray}
Comonotonicity of $(X_1^*,\cdots, X_n^*)\in{\cal{R}}(F_1, \cdots, F_n)$ implies that  (cf. Dhaene et al. (2006))
\begin{eqnarray}
\rho_{g}[X_1]+\cdots+\rho_{g}[X_n]=\rho_{g}[X_1^* + \cdots+X_n^*].
\end{eqnarray}
 Therefore, combining (2.1) with (2.2)  one has
$$\rho_{g}[X_1 + \cdots+X_n]\le \rho_{g}[X_1^*+\cdots+X_n^*],$$
and the desired result follows from Lemma 2.1.

To prove the other implication, we  assume that $(X_1^*,\cdots, X_n^*)\in{\cal{R}}(F_1, \cdots, F_n)$ and
$$X_1+\cdots+X_n\le_{cx} X_1^*+\cdots+X_n^* \,\, {\rm for\ all}\,\, (X_1,\cdots, X_n)\in {\cal{R}}(F_1, \cdots, F_n).$$
From Lemma 2.1 we have that
$$\rho_{g}[X_1 + \cdots+X_n]\le \rho_{g}[X_1^*+\cdots+X_n^*],$$
for all concave distortion functions $g$. In particular,
\begin{eqnarray}
\rho_{g}[X_1^c + \cdots+X_n^c]\le \rho_{g}[X_1^*+\cdots+X_n^*],
\end{eqnarray}
where $(X_1^c,\cdots,X_n^c)$ is the comonotonic counterpart of $(X_1,\cdots,X_n)$.
On the other hand, by Lemma 2.2 we get
\begin{eqnarray}
\rho_{g}[X_1^*+\cdots+X_n^*]\le \rho_{g}[X_1^*]+\cdots+\rho_{g}[X_n^*].
\end{eqnarray}
If  $(X_1^*,\cdots, X_n^*)$ is not  comonotonic, then
\begin{eqnarray*}
\rho_{g}[X_1^*+\cdots+X_n^*]\neq\rho_{g}[X_1^*]+\cdots+\rho_{g}[X_n^*],
\end{eqnarray*}
which, together with (2.4) leads to
\begin{eqnarray}
\rho_{g}[X_1^*+\cdots+X_n^*]< \rho_{g}[X_1^*]+\cdots+\rho_{g}[X_n^*].
\end{eqnarray}
It follows from (2.3) and (2.5) and note  that
$$\rho_{g}[X_1^c +\cdots+ X_n^c]= \rho_{g}[X_1^c] +\cdots+\rho_{g}[X_n^c],$$
we have
\begin{eqnarray*}
 \rho_{g}[X_1^c]+\cdots+\rho_{g}[X_n^c]< \rho_{g}[X_1^*]+\cdots+\rho_{g}[X_n^*],
\end{eqnarray*}
which is obviously a contradiction since
\begin{eqnarray*}
 \rho_{g}[X_1^c]+\cdots+\rho_{g}[X_n^c]=\rho_{g}[X_1^*]+\cdots+\rho_{g}[X_n^*].
\end{eqnarray*}
 Thus, $(X_1^*,\cdots, X_n^*)$ is comonotonic. This ends the proof of Theorem 1.1.

\vskip 0.2cm
 \section{A new proof of Theorem 1.2}
\setcounter{equation}{0}

To prove  Theorem 1.2, we need two useful
lemmas. Lemma 3.1 gives a necessary and sufficient condition for convex order of two rvs which  was given in Proposition 3.4.3  of  Denuit et al. (2005).
\begin{lemma} (Denuit et al. (2005, Proposition 3.4.3)) \  Given two rvs $X$ and $Y$,  then the following statements are equivalent:
\item (1)\ $ X\le_{cx} Y$.
\item (2)\ $E[v(X)]\le E[v(Y)]$   for all convex functions $v$ such that the expectations exist.
\item (3)\   $E[v(X)]\le E[v(Y)]$   for all functions $v$ with $v''\ge 0$ such that the expectations exist.
\end{lemma}
The following lemma, due to Cheung and Lo (2013), will play a crucial role in the proof of  Theorem 1.2.
\begin{lemma}  (Cheung and Lo (2013, Theorem 3.1))\ Let $X_1,\cdots,  X_n$ be non-negative random variables and
$f$ be a convex function such that $E[f(\sum_{i=1}^n X_i)]$ exists.
\item (i) We have
$$E[f(\sum_{i=1}^n X_i)]\ge \sum_{i=1}^n E[f(X_i)]-(n-1)f(0);$$
\item (ii)  if $f$ is strictly convex, then
$$E[f(\sum_{i=1}^n X_i)]= \sum_{i=1}^n E[f(X_i)]-(n-1)f(0)$$
 if and only if $X_1,\cdots,  X_n$  are mutually exclusive random variables in the sense of Dhaene and Denuit (1999).
\end{lemma}
\begin{remark} We remark that the ``if part" is still true when the function $f$ is convex, but
not necessarily strictly convex.
\end{remark}
{\bf Proof of Theorem 1.2}.\  To prove Theorem 1.2, as in the proof to Lemma 3.6 in Cheung and Lo (2014), there are  three cases to consider.

Case 1. $l_1=\cdots=l_n=0$. First we assume $(X_1^*,\cdots, X_n^*)\in{\cal{R}}(F_1, \cdots, F_n)$ is  mutually exclusive. For any convex function $u$ and $(X_1,\cdots, X_n)\in {\cal{R}}(F_1, \cdots, F_n)$,
 By Lemma 3.2 (i)   we have
\begin{eqnarray}
E[u(\sum_{i=1}^n X_i)]\ge \sum_{i=1}^n E[u(X_i)]-(n-1)u(0).
\end{eqnarray}
Thanks to  Lemma 3.2 (ii) and Remark 3.1, mutual exclusivity   of $(X_1^*,\cdots, X_n^*)$ implies that
\begin{eqnarray}
E[u(\sum_{i=1}^n X_i^*)]=\sum_{i=1}^n E[u(X_i^*)]-(n-1)u(0).
\end{eqnarray}
 Therefore, combining (3.1) with (3.2), and note that $E[u(X_1^*)]+\cdots+E[u(X_n^*)]=E[u(X_1)]+\cdots+E[u(X_n)]$,  one has
$$
E[u(\sum_{i=1}^n X_i^*)] \le E[u(\sum_{i=1}^n X_i)],$$
from which and Lemma 3.1, we  deduce that
$$ X_1^*+\cdots+X_n^*\le_{cx}  X_1+\cdots+X_n$$
 for all $(X_1, \cdots, X_n) \in  {\cal{R}}(F_1, \cdots, F_n)$.

To prove the other implication, we  assume that $(X_1^*,\cdots, X_n^*)\in{\cal{R}}(F_1, \cdots, F_n)$ and
$$ X_1^*+\cdots+X_n^*\le_{cx} X_1+\cdots+X_n\,\, {\rm for\ all}\,\, (X_1,\cdots, X_n)\in {\cal{R}}(F_1, \cdots, F_n).$$
From Lemma 3.1 we have that
$$E[u(X_1^*+\cdots+X_n^*)]\le E[u(X_1 + \cdots+X_n)]$$
for all convex functions $u$. In particular,
\begin{eqnarray}
 E[u(X_1^*+\cdots+X_n^*)]\le E[u(X_1^M + \cdots+X_n^M)]
\end{eqnarray}
where $(X_1^M,\cdots,X_n^M)$ is the mutually exclusive counterpart of $(X_1,\cdots,X_n)$.
On the other hand, by Lemma 3.2 and Remark 3.1 we get
\begin{eqnarray}
  E[u(X_1^*+\cdots+X_n^*)]\ge E[u(X_1^*)] + \cdots+ E[u(X_n^*))]-(n-1)u(0),
\end{eqnarray}
and
\begin{eqnarray}
  E[u(X_1^M+\cdots+X_n^M)]= E[u(X_1^M)] + \cdots+ E[u(X_n^M))]-(n-1)u(0),
\end{eqnarray}
If  $(X_1^*,\cdots, X_n^*)$ is not  mutually exclusive, then
\begin{eqnarray}
  E[u(X_1^*+\cdots+X_n^*)]\neq E[u(X_1^*)] + \cdots+ E[u(X_n^*))]-(n-1)u(0).
\end{eqnarray}
Combining (3.3)-(3.5) with (3.6) we get
 $$
 E[u(X_1^*)] + \cdots+ E[u(X_n^*))]<E[u(X_1^M)] + \cdots+ E[u(X_n^M))].
 $$
This contradicts that $(X_1^*,\cdots, X_n^*)$ and   $(X_1^M,\cdots,X_n^M)$  having
the as marginal distributions.
 Thus, $(X_1^*,\cdots, X_n^*)$ is   mutually exclusive.

Case 2.  $(X_1^*,\cdots, X_n^*)$ is mutually exclusive from below. For any
  $(X_1, \cdots, X_n) \in {\cal R}(F_1,\cdots, F_n)$,  then  $Z:=X_i-l_i$  are   non-negative  random
variables, $Z^*:=X_i^*-l_i$ are  non-negative mutually exclusive random
variables. Applying the result in   Case 1 we obtain that
$(X_1^*,\cdots, X_n^*) \; {\rm is \;  mutually\;  exclusive\; } \Leftrightarrow (X_1^*-l_1,\cdots, X_n^*-l_n)   \; {\rm is \;  mutually\;  exclusive\;}$
$$\Leftrightarrow \sum_{i=1}^n (X_i^*-l_i)\le_{cx}    \sum_{i=1}^n (X_i-l_i)\Leftrightarrow \sum_{i=1}^n X_i^*-\sum_{i=1}^n l_i \le_{cx}    \sum_{i=1}^n X_i-\sum_{i=1}^n l_i$$
$$\Leftrightarrow \sum_{i=1}^n X_i^* \le_{cx}    \sum_{i=1}^n X_i.$$
Case 3. $(X_1^*,\cdots, X_n^*)$ is mutually exclusive from above. For any
  $(X_1, \cdots, X_n) \in {\cal R}(F_1,\cdots, F_n)$, applying the result in Case 2, we have  $(X_1^*,\cdots, X_n^*)$ is mutually exclusive from above $\Leftrightarrow$ $(-X_1^*,\cdots, -X_n^*)$ is mutually exclusive from below $\Leftrightarrow$  $-\sum_{i=1}^n X_i^* \le_{cx} -\sum_{i=1}^n X_i $ $\Leftrightarrow$  $\sum_{i=1}^n X_i^* \le_{cx} \sum_{i=1}^n X_i $. The proof of Theorem 1.2 is complete now.

\noindent{\bf Acknowledgements.} \ 
The research   was supported by the National
Natural Science Foundation of China (No. 11171179, 11571198) and  the Research
Fund for the Doctoral Program of Higher Education of China (No. 20133705110002).

\end{document}